# Light Drag Effect of Vacuum Tube Versus Light Propagation in Stationary Vacuum Tube with Moving Source and Receiver

*Ruyong Wang[a,\*], Li Zhan[b], Le He[b], Wenyan Zhang[b], Liang Zhang[b]*

[a] *St. Cloud State University, St. Cloud, MN 56301, USA*

[b] *State Key Laboratory of Advanced Optical Communication Systems and Networks, Key Laboratory for Laser Plasmas (Ministry of Education), Department of Physics and Astronomy, Shanghai Jiao Tong University, Shanghai 200240, China*

**Abstract**

We presented a new way to examine the principle of relativity of Special Relativity. According to the principle of relativity, the light dragging by moving media and the light propagation in stationary media with moving source and receiver should be two totally equivalent phenomena. We select a vacuum tube with two glass rods at two ends as the optical media. The length of the middle vacuum cell is $L$ and the thicknesses of the glass rods with refractive index $n$ are $D_1$ and $D_2$. The light drag effect of the moving vacuum tube with speed $v$ is a first-order effect, $\Delta t = 2(n-1)(D_1 + D_2)v/c^2$, which is independent of $L$ because vacuum does not perform a drag effect. Predicted by the principle of relativity, the change of the light propagation time interval with stationary vacuum tube and moving source and receiver must be the same, i.e., $\Delta \tau \equiv \Delta t = 2(n-1)(D_1 + D_2)v/c^2$. However all analyses have shown that the change of the propagation time interval $\Delta \tau$ is caused by the motion of the receiver during the light propagation in the vacuum tube. Thus, the contribution of the glass rods in $\Delta \tau$ is $2n(D_1 + D_2)v/c^2$, not $2(n-1)(D_1 + D_2)v/c^2$ in $\Delta t$. Importantly, the contribution of the vacuum cell in $\Delta \tau$ is $2Lv/c^2$, not zero in $\Delta t$.

Our analyses are solid in optics. The genuine tests of the prediction of the principle of relativity can be conducted by the experiments with two atomic clocks, or the experiments with fiber Sagnac interferometers.

[\*]Corresponding author: ruwang@stcloudstate.edu

## 1. Introduction

In 1980, it was indicated that the principle of relativity of Special Relativity had not been verified by experiments in a system moving relative to the Earth and a new Michelson-Morley experiment in Space Lab was proposed [1]. Popper thought conducting such an experiment to examine the principle of relativity was a good idea [2]. Unfortunately, nothing ever came of it and no experiment has been carried out in a system moving relative to the Earth to date.

The principle of relativity states that in any system of coordinates in uniform translatory motion, the speed of light is a constant $c$. The generalized Sagnac experiments [3,4], also called linear Sagnac experiments, have shown that in an air-core fiber segment of length $l$ and in uniform translatory motion with speed $v$, the travel time difference between two counter-propagating light beams is a first-order effect, $\Delta t = 2vl/c^2$.

In this paper, we presented a new way to examine the principle of relativity. The principle of relativity of Special Relativity states that the physical laws are the same for all the observers in uniform translatory motion regardless of their different motion statuses. Therefore all the phenomena must be the same in these two cases: a moving vacuum tube with stationary source and receiver and a stationary vacuum tube with moving source and receiver. We conducted the analyses based on optics, and therefore, this way is solid in the analyses. We also indicated the possible experiments. Importantly, the experiments are based on the first-order effect. As comparisons, Lorentz contraction and relativistic time dilation are the second-order effect. Obviously these two cases represent two different moving statuses relative to the Earth. The proposed experiments are the first-order experiments examining the principle of relativity in two systems.

## 2. New way of examining the principle of relativity

We present a new way of examining the principle of relativity: moving source and receiver versus moving group of optical media where the source and receiver always move together with the same speed as shown in **Fig. 1**. Therefore from the viewpoint of relative motion, there are only two entities: a source and a receiver versus a group of optical media, and we investigate two cases: moving group of media with stationary source and receiver (**Fig. 1**a); stationary group of media with moving source and co-moving receiver (**Fig. 1**b). It is well known that the former is the light drag effect. The latter is the light propagation in stationary media with moving source and receiver. According to the principle of relativity of Special Relativity, all the physical phenomena, including the light propagation time interval, are the same in these two cases. In fact, if we have Observers 1 on the group of optical media and Observer 2 on the receiver, for these two observers in uniform translatory motion, they won't be able to identify the difference between these two cases.

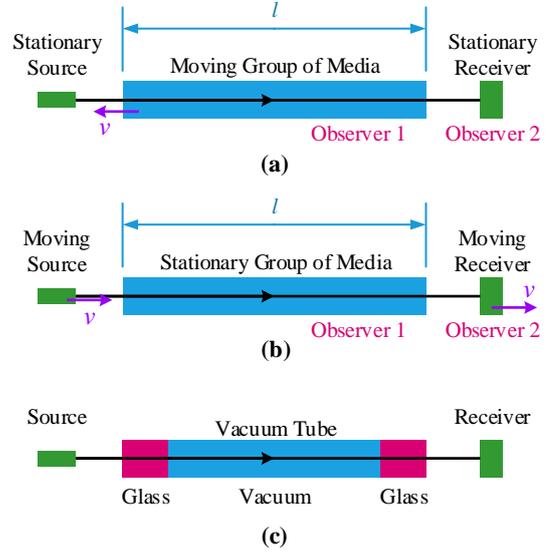

**Fig. 1.** Moving group of optical media versus moving source and co-moving receiver. The group of optical media - a vacuum tube with two glass rods.

The light drag effect of moving media has been an important topic in optics for many years. However, according to the principle of relativity, this effect can be treated as a different problem: light propagation in stationary media with moving source and receiver. Is it really true?

Apparently the previous statements should be true for any optical media. Here we select a vacuum tube, a tube having two glass rods with very low dispersion as two ends and the air is extracted as shown in **Fig. 1**c.

## 3. Primary analysis

The sizes of the vacuum tube are shown in **Fig. 2**a. The lengths of the two glass rods are $D_1$ and $D_2$ and the length of the vacuum cell is $L$. The whole tube is called a vacuum tube and the part without two glass rods is the vacuum cell.

The propagation time interval for a light beam which starts from a stationary source, passes the stationary vacuum tube, then arrives the stationary receiver is denoted as $t_0$ (**Fig. 2**a). Accordingly, the time interval in **Fig. 2**b is $t$ and the time interval in **Fig. 2**c is $\tau$.

In all these cases, the time interval of a light beam propagating from the source to the receiver always consists of three periods: before entering the vacuum tube, inside the vacuum tube, and after leaving the vacuum tube.

For the case in **Fig. 2**a, we have

$$t_0 = t_{01} + t_{02} + t_{03}$$
$$= l_L/c + (nD_1/c + L/c + nD_2/c) + l_R/c \quad (1)$$

where $n$ is the refractive index of the glass.

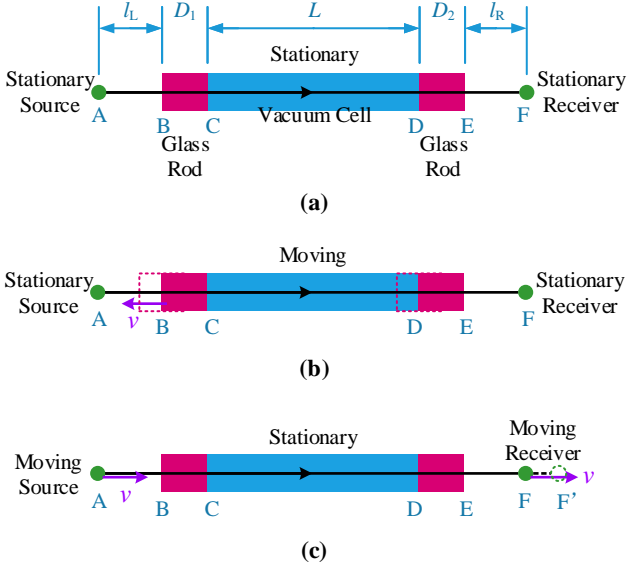

**Fig. 2.** Moving vacuum tube vs stationary vacuum tube.

Let us examine two cases in **Fig. 2**b and in **Fig. 2**c.

First the case in **Fig. 2**b. It is the case of light dragging by optical media and it has been fully investigated. Compared with the stationary vacuum tube, the drag effect in this case [5] is

$$\Delta t = t - t_0 = (n-1)(D_1 + D_2)v/c^2 \quad (2)$$

Obviously only the glass rods contribute to the drag effect while the vacuum cell does not contribute anything to the drag effects. Here we neglect the dispersion term. Actually if including the dispersion term, we have $\Delta t = (n - 1 + \omega dn/d\omega)(D_1 + D_2)v/c^2$ [6-9], and the amount contributed by the dispersion term is small even for dispersive glasses.

Then the case in **Fig. 2**c. Based on the principle of relativity of Special Relativity, this case is totally equivalent to the case in **Fig. 2**b because as we mentioned above, for each observer of the two observers 1 and 2, these two cases are the same. Therefore, we must have

$$\Delta \tau = \tau - t_0 \equiv \Delta t = (n-1)(D_1 + D_2)v/c^2 \quad (3)$$

Now let us examine whether this prediction of the principle of relativity is true. The whole travel time difference $\Delta \tau$ must exist in three time periods, i.e., $\Delta \tau = \Delta \tau_1 + \Delta \tau_2 + \Delta \tau_3$.

For the first period $\tau_1$ before entering the stationary vacuum tube, because the speed of light is independent of the motion of the source, we have

$$\Delta \tau_1 = \tau_1 - t_{01} = 0 \quad (4)$$

For the second period $\tau_2$ inside the stationary vacuum tube, although the motion of the source causes a Doppler effect, the change of the frequency of incoming light beam in the vacuum tube, the dispersion of the glass rods is very low, so the change of the frequency does not cause any noticeable change of the propagation time interval in the vacuum tube. Hence we have

$$\Delta \tau_2 = \tau_2 - t_{02} = 0 \quad (5)$$

Therefore the travel time difference $\Delta \tau$ predicted by the principle of relativity should exist only in the third period $\tau_3$, that is $\Delta \tau_3 = \tau_3 - t_{03} = \Delta \tau$ $= (n-1)(D_1 + D_2)v/c^2$.

As shown in **Fig. 2**c, when the light beam exits from the right end of the vacuum tube E, the receiver has moved from F to F' and $\Delta_{FF'} = v(\tau_1 + \tau_2) = l_L v/c + n(D_1 + D_2)v/c + Lv/c$. Clearly the light beam has to spend more time to catch the moving receiver, and the time increase is $\Delta_{FF'}/c = l_L v/c^2 + n(D_1 + D_2)v/c^2 + Lv/c^2$. If we list the predicted $\Delta \tau_3 = (n-1)(D_1 + D_2)v/c^2$ here, we can find two differences between these two results. Firstly, the existence of the two glass rods causes a time increase of $n(D_1 + D_2)v/c^2$, how does this match with $\Delta \tau_3 = (n-1)(D_1 + D_2)v/c^2$? Where does the factor $(n-1)$ come from in this case? (We will further discuss this problem in Appendix.) Secondly and more seriously, the existence of the vacuum cell causes a time increase of $Lv/c^2$. However, required by the principle of relativity, this time increase does not contribute any time difference for $\Delta \tau_3$. Actually if the principle of relativity of Special Relativity is true in this scenario, the light beam in the vacuum tube must have a very strange and selective behavior: when light beam passes the glass rods, no matter how short the rods are, 1 m, 1 dm, or even 1 cm, 1 mm, they will contribute a finite $\Delta \tau_g$. However when the light beam passes the vacuum cell, no matter how long the vacuum cell is, 1 m, 10 m, or even 100 m, 1000 m, the vacuum cell will not contribute any finite $\Delta \tau$. That is, $\Delta \tau_{VC} \equiv 0$ is required.

In summary, the prediction of the principle of relativity $\Delta \tau \equiv \Delta t = (n-1)(D_1 + D_2)v/c^2$ is unlikely true.

## 4. Analysis for specific cases

It would be very interesting to analyze the prediction of the principle of relativity with two vacuum tubes as shown in **Fig. 3**. Tube A has glass rods with length $D_A$ and a vacuum cell with length $L_A$. Tube B has glass rods with length $D_B$ and a vacuum cell with length $L_B$. And we choose $D_A=100D_B$ (e.g., $D_A=10$ cm and $D_B=1$ mm) and $L_A=0.01L_B$ (e.g., $L_A=10$ cm and $L_B=10$ m).

Let us first conduct the light drag experiments (**Fig. 3**a). Clearly we will have the following results.

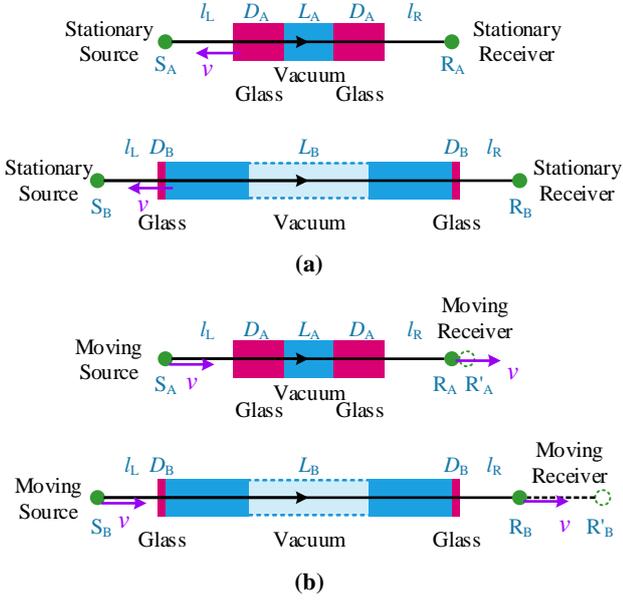

**Fig. 3.** Two vacuum tubes with different lengths of the glass rods and vacuum cells, $D_A = 100D_B$ and $L_A = 0.01L_B$.

Comparing with the case where all parts are stationary, we have drag effects

$$\Delta t_A = 2(n-1)D_A v/c^2$$
$$\Delta t_B = 2(n-1)D_B v/c^2 \quad (6)$$

and $\Delta t_A = 100\Delta t_B$, because $D_A = 100D_B$.

Now let us conduct the experiment with stationary vacuum tubes and moving sources and receivers (**Fig. 3**b). According to the principle of relativity, the experimental results should be

$$\Delta\tau_A \equiv \Delta t_A = 2(n-1)D_A v/c^2$$
$$\Delta\tau_B \equiv \Delta t_B = 2(n-1)D_B v/c^2 \quad (7)$$

and $\Delta\tau_A = 100\Delta\tau_B$.

As indicated above, $\Delta\tau$ is entirely contributed by $\Delta\tau_3$, then we have

$$\Delta\tau_{A3} = \Delta\tau_A = 2(n-1)D_A v/c^2$$
$$\Delta\tau_{B3} = \Delta\tau_B = 2(n-1)D_B v/c^2 \quad (8)$$

and $\Delta\tau_{A3} = 100\Delta\tau_{B3}$.

As shown in **Fig. 3**b, the time period that a light beam propagates inside Tube A is much shorter than that inside Tube B. Hence, the moving distance of receiver $R_A$ during this period is much shorter than the moving distance of receiver $R_B$, i.e., $\Delta_{R_A R'_A} \ll \Delta_{R_B R'_B}$. Obviously $\Delta_{R_A R'_A} \ll \Delta_{R_B R'_B}$ contradicts with $\Delta\tau_{A3} = 100\Delta\tau_{B3}$ and the latter is an extremely odd prediction of the principle of relativity.

## 5. Theoretical analysis of the light propagation in the vacuum tube

Now let us analyze the light propagation in the vacuum tube theoretically and we consider the light propagation in two opposite directions (**Fig. 4**).

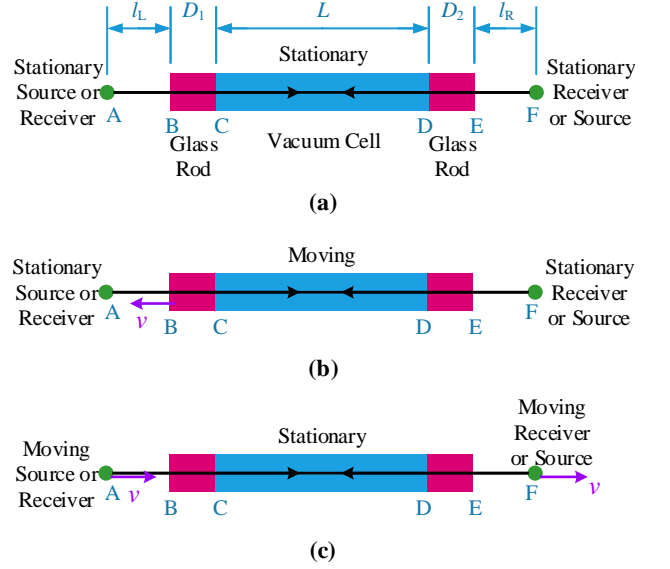

**Fig. 4.** Light propagation in two opposite directions - in moving vacuum tube vs in stationary vacuum tube.

a) First, all the source and receiver and vacuum tube are stationary (**Fig. 4**a).

For a light beam propagating from A to F, we have
$$t_0 = t_{01} + t_{02} + t_{03}$$
$$= l_L/c + (nD_1/c + L/c + nD_2/c) + l_R/c \quad (9)$$

For a light beam from F to A, we have
$$t'_0 = t'_{03} + t'_{02} + t'_{01}$$
$$= l_R/c + (nD_1/c + L/c + nD_2/c) + l_L/c = t_0 \quad (10)$$

b) A and F are stationary and the vacuum tube is moving left with a speed $v$ (**Fig. 4**b). For the light propagation time from A to F, it is a problem of light dragging by moving vacuum tube. It has been investigated theoretically and experimentally and we should have the drag effect of the glass rods and the vacuum cell does not have a drag effect [5]. The difference between the propagation time intervals of two beams in opposite directions is

$$\Delta t = t - t' = 2(n-1)(D_1 + D_2)v/c^2 \quad (11)$$

c) When A and F are co-moving right with a speed $v$ and the vacuum tube is stationary (**Fig. 4**c), for the light propagation time interval from A to F, we have

$$\tau = \tau_1 + \tau_2 + \tau_3 \tag{12}$$

$$\tau_1 = l_L / c \tag{13}$$

because the speed of light is independent of the motion of the source.

$$\tau_2 = nD_1/c + L/c + nD_2/c \tag{14}$$

As mentioned before, the vacuum tube is stationary and the frequency change caused by the Doppler effect of the moving source does not yield a noticeable change to the time interval inside the vacuum tube.

When the light beam leaves the right end of the vacuum tube E, time $(\tau_1 + \tau_2)$ has passed, so F has moved a distance of $v(\tau_1 + \tau_2)$ farther.

Therefore we have

$$\begin{aligned}\tau_3 &= [l_R + v(\tau_1 + \tau_2)]/c \\ &= l_R/c + [l_L + n(D_1+D_2) + L]v/c^2\end{aligned} \tag{15}$$

and

$$\begin{aligned}\tau &= [l_L + n(D_1+D_2) + L + l_R]/c \\ &+ [l_L + n(D_1+D_2) + L]v/c^2\end{aligned} \tag{16}$$

For the light propagation time from F to A, we have

$$\tau' = \tau'_3 + \tau'_2 + \tau'_1 \tag{17}$$

$$\tau'_3 = l_R / c \tag{18}$$

$$\tau'_2 = nD_2/c + L/c + nD_1/c \tag{19}$$

When the light beam leaves the left end of the vacuum tube B, time $(\tau'_3 + \tau'_2)$ has passed, so A has moved a distance of $v(\tau'_3 + \tau'_2)$ closer.

Therefore we have

$$\begin{aligned}\tau'_1 &= [l_L - v(\tau'_3 + \tau'_2)]/c \\ &= l_R/c - [l_R + n(D_1+D_2) + L]v/c^2\end{aligned} \tag{20}$$

and

$$\begin{aligned}\tau' &= [l_R + n(D_1+D_2) + L + l_L]/c \\ &- [l_R + n(D_1+D_2) + L]v/c^2\end{aligned} \tag{21}$$

Hence, the travel time difference between two counter-propagating light beams is

$$\Delta\tau = \tau - \tau' = [l_L + 2n(D_1+D_2) + 2L + l_R]v/c^2 \tag{22}$$

Apparently the analysis shows that $\Delta\tau \neq \Delta t$ and their difference is $\Delta\tau - \Delta t = [l_L + 2(D_1+D_2) + 2L + l_R]v/c^2$. Besides, each part's contribution to the total time difference is also different. The contribution of the glass rods is $2n(D_1+D_2)v/c^2$ in $\Delta\tau$, not $2(n-1)(D_1+D_2)v/c^2$ in $\Delta t$. The contribution of the vacuum cell is $2Lv/c^2$ in $\Delta\tau$, not zero in $\Delta t$.

## 6. The genuine tests of the principle of relativity

For the cases we mentioned above, the principle of relativity of Special Relativity gives an odd and, to put it in Popper's words, a risky prediction [10], $\Delta\tau \equiv \Delta t = 2(n-1)(D_1+D_2)v/c^2$. That is to say, the travel time difference $\Delta\tau$ can neither be zero, nor be related to the length of the vacuum cell, $L$. $\Delta\tau$ can only be related to the lengths of the glass rods, $D_1$ and $D_2$, and the factor has to be $(n-1)$. Therefore the experiments examining whether the prediction is true or not will be the genuine tests of the principle of relativity.

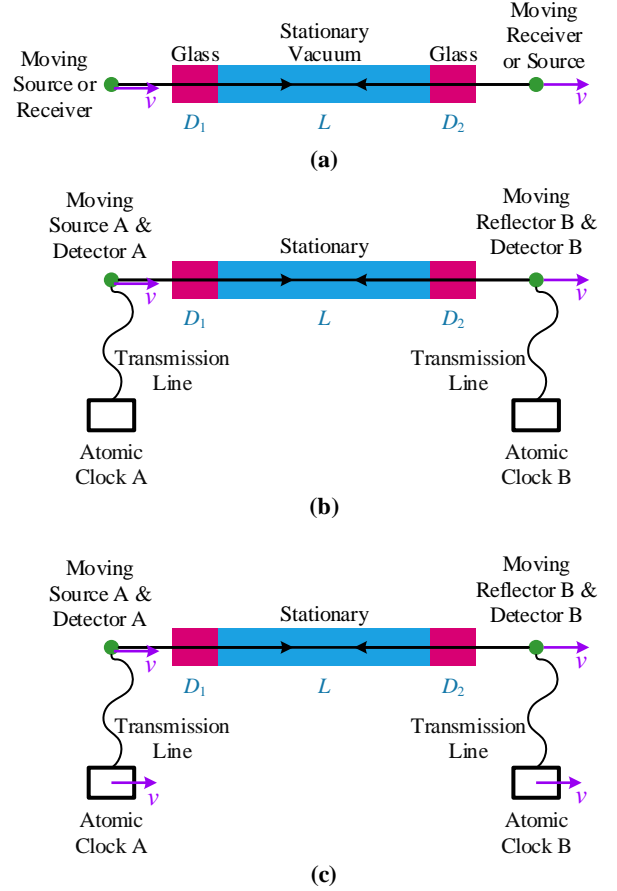

**Fig. 5.** Clock version of the genuine tests.

There are two ways of conducting the genuine tests. First the clock version of the experiments is shown as **Fig. 5**b and the light path mentioned above is also placed there as **Fig. 5**a. Clearly the light path in **Fig. 5**b matches with the light path needed for the experiment.

When $v = 0$, for a light beam, we record the time leaving A with Clock A as $t_0(A)$ and the time arriving B with Clock B as $t_0(B)$; when it reflects back from B, record the time arriving A with Clock A as $t'_0(A)$ and we have

$$[t'_0(A) - t_0(B)] - [t_0(B) - t_0(A)] = \Delta T \tag{23}$$

$\Delta T$ can be positive or negative or zero because two clocks

are not synchronized. Now we set $\Delta T = 0$ to synchronize two clocks. That is, the two clocks are synchronized if the recordings of clock B are added by $\Delta T/2$.

When the speed is $v$, do these again: record the time leaving A with Clock A as $\tau(A)$ and the time arriving B with Clock B as $\tau(B)$ when it reflects back, record the time arriving A with Clock A as $\tau'(A)$ and we have

$$\Delta \tau = [\tau'(A) - \tau(B)] - [\tau(B) - \tau(A)] \quad (24)$$

Predicted by the principle of the relativity, the result must be $\Delta \tau \equiv 2(n-1)(D_1 + D_2)v/c^2$.

In **Fig. 5**b, two atomic clocks are stationary, so they are always synchronized. The two transmission lines are deforming. However their deformation does not cause a net effect because they are relatively short and they have the same deformation, A better, but more difficult configuration is shown in **Fig. 5**c where two atomic clocks are co-moving with speed $v$. In this case, the two clocks are still synchronized because their speeds are the same. Therefore it is expected two configurations in **Fig. 5**b and **Fig. 5**c yield the same results.

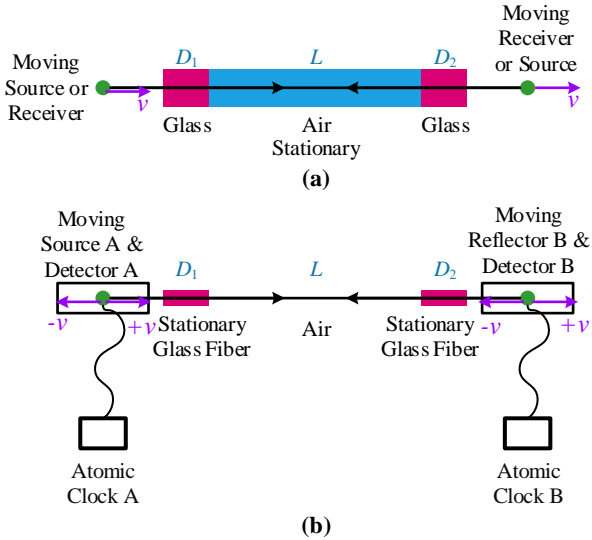

**Fig. 6.** Clock version of the genuine tests with two spaced stationary fibers.

Because the refractive index of the air is very close to 1, we can use a glass tube without extracting the air as shown in **Fig. 6**a. Its clock version is shown as **Fig. 6**b there two stationary glass fibers are used to replace the glass rods, so their lengths can be very long and the distance between the two spaced fibers can be very large as well. The experiment can be conducted with two cases of speeds $+v$ and $-v$, and the final result $\Delta \tau$ will be doubled if we compare those two cases.

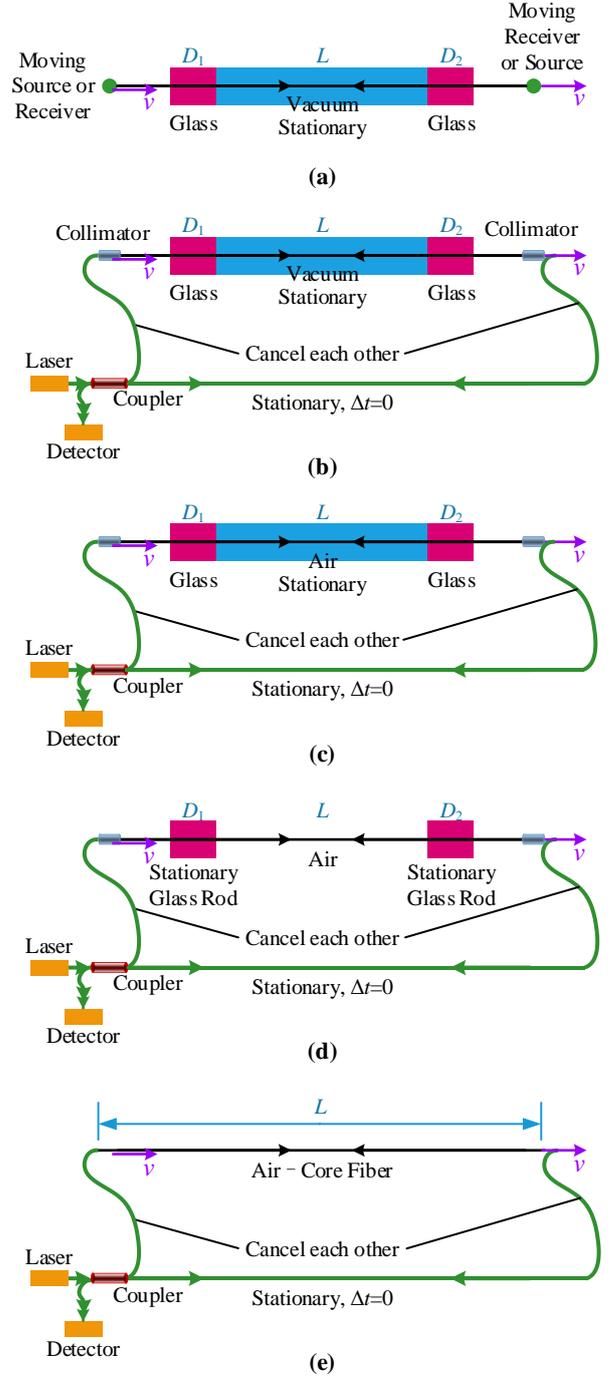

**Fig. 7.** Experiments of stationary media with moving source and receiver using a fiber Sagnac interferometer.

For the interferometric experiment, we place the light path of the genuine test of the principle of relativity again in **Fig. 7**a. When we conduct an interferometric experiment, the light paths of an interferometer constitute a loop so its light paths must be more than the light path needed. Therefore, similarly to the linear Sagnac experiments [4] where a fiber parallelogram is used, we can build a fiber parallelogram with two collimators as shown in **Fig. 7**b for the genuine test. In the fiber parallelogram, the top arm moves with a constant speed $v$ and the bottom arm is stationary. While moving, the two relatively short side-arms are kept the same shape so that the travel time differences in these two side-arms cancel each other and there is no travel time difference in the bottom stationary arm. Therefore, the detected travel time difference of the interferometer is contributed solely by the motion of the top arm, and light path of the top arm in **Fig. 7**b matches with the light path needed for the experiment.

Because the refractive index of the air is very close to unity, we can use a glass tube without extracting the air as shown in **Fig. 7**c and therefore, the configuration in **Fig. 7**d is also the same.

It is not difficult to check whether the experimental result is $\Delta \tau = 2(n-1)(D_1 + D_2)v/c^2$ exactly as predicted by the principle of relativity. As comparisons, the analysis above gives the result $\Delta \tau = [l_L + 2n(D_1 + D_2) + 2L + l_R]v/c^2$ and the parallelogram experiment of the linear Sagnac effect with an air-core fiber segment [4] is shown in **Fig. 7**c and the experimental result is $\Delta t = 2Lv/c^2$.

Based on these, it is expected that the prediction of the principle of relativity, $\Delta \tau = 2(n-1)(D_1 + D_2)v/c^2$, most likely cannot pass the genuine tests.

## 7. Conclusions

To summarize, we have presented a new way of examining the principle of relativity of Special Relativity. We select a vacuum tube with two glass rods at two ends as the optical media. The drag effect of the moving vacuum tube is $\Delta t = 2(n-1)(D_1 + D_2)v/c^2$, which is independent of the length of the vacuum cell. Predicted by the principle of relativity, the change of the light propagation time interval with stationary vacuum tube and moving source and receiver should be the same. Our analyses show that the change of the propagation time interval $\Delta \tau$ is caused by the motion of the receiver during the propagation of the light beam in the vacuum tube. The contribution of the glass rods in $\Delta \tau$ is $2n(D_1 + D_2)v/c^2$, not $2(n-1)(D_1 + D_2)v/c^2$. More importantly, the contribution of the vacuum cell in $\Delta \tau$ is $2Lv/c^2$, not zero. The genuine tests of the prediction can be conducted with the experiments with two atomic clocks, or the experiments with fiber Sagnac interferometers.

## Appendix

For a non-dispersive medium, we investigate its drag effect and the light propagation in stationary medium with moving source and co-moving receiver.

In **Fig. 8**, the time interval of a light beam propagating from S to R always consists of three periods: before entering the medium, inside the medium, and after leaving the medium.

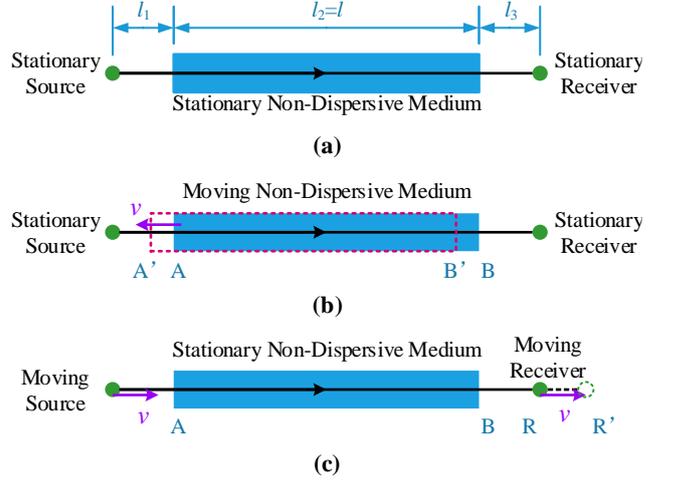

**Fig. 8.** Moving non-dispersive medium versus moving source and receiver.

When all the medium, source and receiver are stationary (**Fig. 8**a), we have

$$t_0 = t_{01} + t_{02} + t_{03} = l_1/c + nl/c + l_3/c \quad (A.1)$$

For the drag effect of the moving medium as shown in **Fig. 8**b, we have

$$t = t_1 + t_2 + t_3 \quad (A.2)$$

The drag effect of a non-dispersive medium is $\Delta t = t - t_0 = (n-1)vl/c^2$. That is, there is a factor $(n-1)$ in $vl/c^2$. Obviously it is correct because the drag effect of vacuum $(n=1)$ is zero. Let us find where $\Delta t = (n-1)vl/c^2$ comes from.

Firstly, because the medium is moving left, the propagation time interval inside the medium becomes shorter [4, 5] and the difference is

$$\Delta t_2 = t_2 - t_{02} = -vl/c^2 \quad (A.3)$$

Secondly when the light beam left the right end of the medium, B has moved to B' and the moving distance is $vt_2 = v(nl/c - vl/c^2)$. Therefore the light beam will spend more time to arrive the stationary receiver and the difference is

$$\Delta t_3 = vt_2/c = v(nl/c - vl/c^2)/c \approx nvl/c^2 \quad (A.4)$$

Hence we have the total difference of the propagation time interval

$$\Delta t = \Delta t_2 + \Delta t_3 = (n-1)vl/c^2 \quad (A.5)$$

For the light propagation in stationary non-dispersive medium with moving source and receiver (**Fig. 8**c), we have

$$\tau = \tau_1 + \tau_2 + \tau_3 \quad (A.6)$$

Because the speed of light is independent of the motion of the source and the non-dispersive medium is stationary, we have

$$\Delta\tau_1 = 0, \quad \tau_1 = t_{01} = l_1/c$$
$$\Delta\tau_2 = 0, \quad \tau_2 = t_{02} = nl/c \tag{A.7}$$

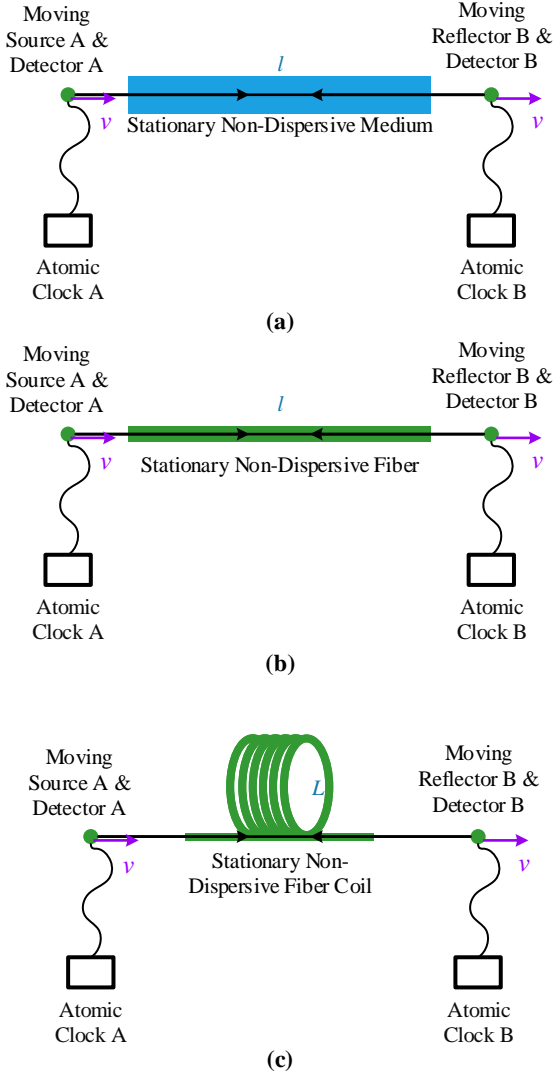

Fig. 9. Clock version of the experiment.

The principle of relativity of Special Relativity in this case requires $\Delta\tau_3 = \Delta\tau \equiv \Delta t = (n-1)vl/c^2$. Could this possibly be true?

Actually when the light beam leaves the right end of medium B, the receiver has moved to R' and the moving distance is $v(\tau_1 + \tau_2) = v(l_1/c + nl/c)$.

Therefore the light beam will spend more time to catch the moving receiver and the difference is

$$\Delta\tau_3 = v(\tau_1 + \tau_2)/c = vl_1/c^2 + nvl/c^2 \tag{A.8}$$

Generally, $l_1$ is much shorter than $l$, and we have

$$\Delta\tau = \Delta\tau_3 = nvl/c^2 \tag{A.9}$$

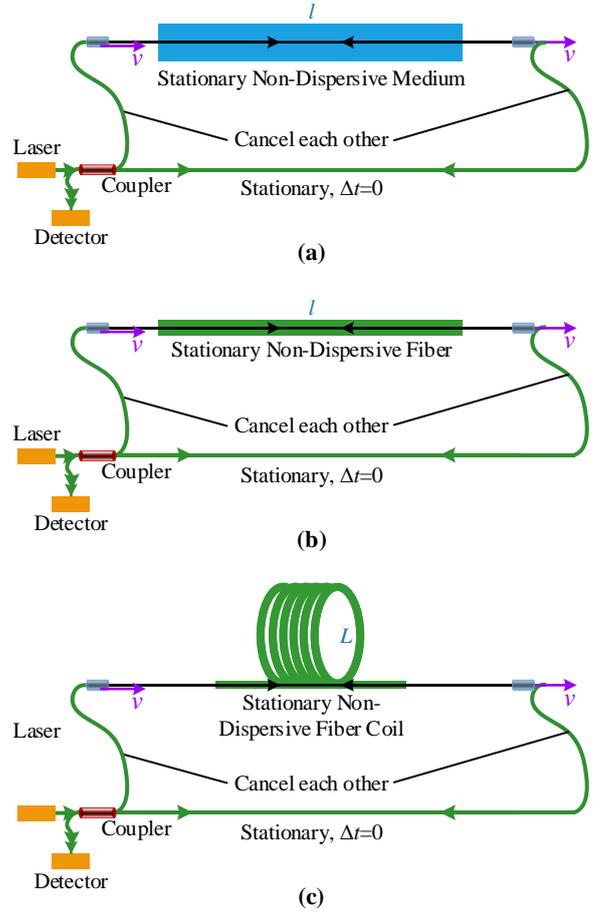

Fig. 10. Experiments of stationary non-dispersive medium with moving source and receiver using a fiber Sagnac interferometer.

In conclusion, we find not only $\Delta\tau_2 \ne \Delta t_2$, but also $\Delta\tau \ne \Delta t$. Contrary to what the principle of relativity predicts, we have

$$\Delta t = (n-1)vl/c^2$$
$$\Delta\tau = nvl/c^2 \tag{A.10}$$

It is clear that the difference between them is that one has a factor $(n-1)$, and the other has a factor $n$.

To examine whether the factor is $n$ or $(n-1)$ for $\Delta\tau$, we can conduct the experiments as shown in **Fig. 9**. Fig. 9a is the clock version of the experiment. In **Fig. 9** b, a non-dispersive fiber is used so the length $l$ can be much longer. Because in these configurations the non-dispersive medium is stationary and the total propagation time interval inside the medium is proportional to the length of the medium, we can greatly increase the total time interval by using a non-dispersive fiber coil with total length $L$ as shown in **Fig. 9**c. Apparently in this case $\Delta\tau = nvL/c^2$ and it is not difficult to examine whether it is really the case.

Experiments in **Fig. 10** are the interferometric counterparts of the experiments in **Fig. 9**.